\documentclass[aps,pre,preprint]{revtex4}

\usepackage{epsfig}

\begin{document}

\title{Nonlinear mirror modes in the presence of hot electrons}
\author{E.A. Kuznetsov$^{(a,b)}$\/\thanks{e-mail:kuznetso@itp.ac.ru}, 
T. Passot $^{(c)}$ and P.L. Sulem $^{(c)}$}
\affiliation{\small \textit{$^{(a)}$P.N. Lebedev Physical Institute RAS, 53 Leninsky
Ave., 119991 Moscow, Russia\\
$^{(b)}$Space Research Institute RAS, 84/32 Profsoyuznaya
 str., 117997, Moscow, Russia\\
$^{(c)}$Universit\'e de Nice-Sophia Antipolis, CNRS,
Observatoire de la C\^ote d'Azur, PB 4229,
06304 Nice Cedex 4, France}}

\begin{abstract}

A non-perturbative calculation of the gyrotropic pressures associated with large-scale 
mirror modes is performed, taking into account a finite,  possibly anisotropic 
electron temperature. In the small-amplitude limit, this leads to an extension of 
an asymptotic model  previously derived for  cold electrons. 
A model equation for the profile of subcritical finite-amplitude large-scale structures 
is also presented. 

\end{abstract}

\maketitle


\section{Introduction}

Pressure-balanced magnetic structures in the form of strong magnetic
enhancements (humps) and depressions (holes) that are quasi-stationary in
the plasma frame, with no or little change in the magnetic field direction,
are commonly observed in regions of the solar wind and of planetary
magnetosheaths with relatively large $\beta $ and a dominant (generally
ion) temperature in the transverse direction ( see, for instance, \cite{S11, Genot09b} and
references therein). The origin of these structures is still not fully
understood, but they are usually viewed as nonlinearly saturated states of the
mirror instability (MI) discovered by Vedenov and Sagdeev 
\cite{VedenovSagdeev}. It is a kinetic instability whose growth rate
was first obtained under the assumption of cold electrons, a regime where
the contributions of the parallel electric field $E_{\Vert }$ can be
neglected. However, in realistic space plasmas, the electron temperature can
hardly be ignored \cite{Stverak08}. The linear theory retaining the electron
temperature and its possible anisotropy, in the quasi-hydrodynamic limit
(which neglects finite Larmor radius corrections), was developed in the case
of bi-Maxwellian distribution functions by several authors 
\cite{Stix62}--\cite{Hell07}. A general estimate of the growth rate 
under the sole condition that it is small compared with the ion
gyrofrequency (a condition reflecting  close vicinity to threshold) is
presented in \cite{KPS12a}. The instability then develops in
quasi-perpendicular directions, making the parallel magnetic perturbation
dominant. This analysis includes in particular regimes with a significant
electron temperature anisotropy for which the instability extends beyond the
ion Larmor radius. In the limit where the instability is limited to scales
large compared with the ion Larmor radius, only the leading order
contribution in terms of the small parameter $\gamma /(|k|_{z}v_{\|i})$ is
to be retained in estimating Landau damping, and the growth rate is given by 
\begin{eqnarray}
&&\gamma =\frac{2}{\sqrt{\pi }}\frac{T_{\Vert i}}{T_{\perp i}}\frac{%
|k_{z}|v_{\Vert i}}{E}\Big \{\Gamma  -\frac{1}{\beta _{\perp }}\Big (1+\frac{\beta _{\perp }-\beta
_{\Vert }}{2}\Big )\frac{k_{z}^{2}}{k_{\perp }^{2}}  \nonumber \\
&&\qquad -\frac{3}{4(1+\theta _{\perp })}\Big (\frac{T_{\perp i}}{T_{\Vert i}%
}-1\Big )(1+F)k_{\perp }^{2}r_{L}^{2}\Big \},  \label{growth_rate}
\end{eqnarray}%
where 
\begin{equation}
\Gamma =\frac{T_{\perp i}}{T_{\parallel i}}\frac{(\theta _{\parallel
}+\theta _{\perp })^{2}+2\theta _{\parallel }(\theta _{\perp }^{2}+1)}{%
2\theta _{\parallel }(1+\theta _{\perp })(\theta _{\parallel }+1)}-1-\frac{1%
}{\beta _{\perp }}  \label{newthreshold}
\end{equation}%
measures the distance to threshold and 
\begin{eqnarray}
E &=&\frac{1+\theta _{\perp }}{(1+\theta _{\parallel })^{2}}\left[ 2+\theta _{\perp
}(4+\theta _{\perp })+\theta _{\parallel }^{2}\right]  \nonumber \\
F &=&\frac{T_{\Vert e}}{T_{\Vert e}+T_{\Vert i}}\Big \{-1+\frac{\theta
_{\perp }}{\theta _{\Vert }}  \nonumber \\
&&-\frac{2}{3}\frac{T_{\Vert i}}{T_{\perp i}}\Big [\Big (\frac{T_{\Vert i}}{%
T_{\perp i}}-1\Big )\frac{1}{\beta _{\perp i}}-\theta _{\perp }\Big (\frac{%
T_{\perp e}}{T_{\Vert e}}-1\Big )\Big ]\Big \}.  \nonumber
\end{eqnarray}%
Here, ${T_{\perp \alpha }}$ and ${T_{\Vert \alpha }}$ are the perpendicular
and parallel (relative to the ambient magnetic field $\mathbf{B}_{0}$ 
taken in the $z$ direction)
temperatures of the species $\alpha $ ($\alpha =i$ for ions and $\alpha =e$
for electrons ), $\theta _{\perp }={T_{\perp e}}/{T_{\perp i}}$, $\theta
_{\Vert }={T_{\Vert e}}/{T_{\Vert i}}$ and $\beta _{\perp }=\beta _{\perp
i}+\beta _{\perp e}$ with $\beta _{\perp \alpha }=8\pi p{_{\perp \alpha }/B}%
_{0}^{2}$ where $p{_{\perp \alpha }}$ is the perpendicular thermal pressure
(similar definition for $\beta _{\Vert }$). Furthermore, the parallel
thermal velocity is defined as $v_{\Vert \alpha}=\sqrt{{2T_{\Vert \alpha}}/{m_{\alpha}}}$,
and $r_{L}=({2{T_{\perp i}/m_{p})^{1/2}/\Omega _{i}}}
$ denotes the ion Larmor radius ($\Omega _{i}=eB_{0}/m_{i}c$ is the ion
gyrofrequency).

The growth rate given by Eq. (\ref{growth_rate}) has the same structure 
as in the cold electron regime considered  
in \cite{Hall79} in the case of bi-Maxwellian ions 
and generalized in \cite{Pokho05} and \cite
{Hell07} to an arbitrary distribution function. The first term within the
curly brackets  provides the threshold condition which 
coincides with that 
given in 
\cite{Stix62}--\cite{Hall79}. The second one reflects the magnetic field
line elasticity and the third one (where $F$ depends on the electron
temperatures due to  the coupling between the species induced by the
parallel electric field which is relevant for hot electrons)
provides the arrest of the instability at small scales by finite Larmor
radius (FLR) effects.

An aim of this letter is to extend to hot electrons the weakly nonlinear
analysis  previously developed for cold electrons \cite{KPS07a,KPS07b}. 
Since in this asymptotics, FLR contributions appear only at the linear
level, the idea is to use the drift kinetic formalism
to calculate the nonlinear terms. We show that the equation governing
the evolution of  weakly nonlinear mirror modes has
the same form as in the case of cold electrons. In particular, the sign of the
nonlinear coupling coefficient that prescribes the shape of
mirror structures, is not changed. This equation 
is of gradient type equation with a  free
energy (or a Lyapunov functional) which is unbounded  from below.
This leads to  finite-time blowing-up solutions \cite{ZK12}, associated with 
the existence of a subcritical bifurcation 
\cite{KPS07a,KPS07b}. To describe subcritical stationary mirror
structures in the strongly nonlinear regime, we present an anisotropic MHD model 
where the perpendicular and parallel pressures are determined
from the drift kinetic equations in the adiabatic
approximation, in the form of prescribed functions of the magnetic field amplitude.

\section{Basic equations}

A main condition characterizing mirror modes,  at least near threshold,  is
provided by the force balance equation 
\begin{eqnarray}
&&-\nabla \Big(p_{\perp }+\frac{B^{2}}{8\pi }\Big)+\Big[1+ \frac{4\pi }{B^{2}%
}(p_{\perp }-p_{\Vert })\Big]\frac{(\mathbf{B}\cdot \nabla) \mathbf{B}}{4\pi 
}  \nonumber \\
&&+ \mathbf{B}(\mathbf{B}\cdot\nabla) \Big (\frac{p_\perp -p_\|}{B^2} \Big ) %
-\nabla \cdot \mathbf{\Pi}=0,  \label{balance-nl}
\end{eqnarray}
where the pressure tensor, viewed as the  the sum of the contributions of the various species,
has been written as the sum of a gyrotropic part
characterized by the parallel ($p_{\Vert }=\sum_{\alpha}p_{\Vert \alpha }$) and perpendicular 
($p_{\perp }=\sum_{\alpha }p_{\perp \alpha }$) pressures, and 
of a gyroviscous contribution $\mbox{\boldmath $\Pi$}$
originating from the sole ion FLR effects when concentrating on scales large compared with the
electron Larmor radius.  As mentioned above,
FLR effects  arising only at the linear level with
respect to the amplitude of the perturbations, the other linear and nonlinear 
contributions can be evaluated from the drift kinetic equation for each particle species
\begin{equation}
\frac{\partial f_{\alpha }}{\partial t}+v_{\Vert }\mathbf{b}\cdot \nabla
f_{\alpha }+\Big[-\mu \mathbf{b}\cdot \nabla B+\frac{e_{\alpha }}{m_{\alpha
} }E_{\Vert }\Big]\frac{\partial f_{\alpha }}{\partial v_{\Vert }}=0
\label{mainkin}
\end{equation}%

We ignore the transverse electric drift which is subdominant for mirror
modes. In this approximation, both ions and electrons move in the direction
of the magnetic field (defined by the unit vector $\mathbf{b}=\mathbf{B}/B$)
under the effect of the magnetic force $\mu \mathbf{\ b}\cdot\nabla B$ and
the parallel electric field $E_{\Vert }=-\mathbf{b}\cdot \nabla\phi$ where
the magnetic moment $\mu=v_{\perp }^{2}/(2B)$ is an adiabatic invariant
which plays the role of a parameter in Eq. (\ref{mainkin}). Here $\phi$ is
the electric potential. The quasi-neutrality condition $n_{e}=n_{i}\equiv n$%
, where $n_{\alpha }=B\int f_{\alpha }d\mu dv_{\Vert }d\varphi \equiv \int
f_{\alpha}d^{3}v$, is used to close the system and eliminate $E_\|$.

In this framework where FLR effects are neglected, the gyrotropic pressures 
 are given  by 
$p_{\alpha \Vert }\equiv m_{\alpha }\int v_{\Vert }^{2}f_{\alpha }d^{3}v
=m_{\alpha }B\int v_{\Vert }^{2}f_{\alpha }d\mu dv_{\Vert}d\varphi $,  
and $p_{\alpha \perp } \equiv \frac{1}{2}m_{\alpha }\int v_{\perp }^{2}f_{\alpha }d^{3}v
=m_{\alpha }B^{2}\int \mu f_{\alpha }d\mu dv_{\Vert}d\varphi$.

The asymptotic equation governing the mirror dynamics near threshold is obtained by
expanding Eqs. (\ref{balance-nl}), (\ref{mainkin}) and the quasi-neutrality
condition, with  the pressure tensor elements for each species
computed near a bi-Maxwellian equilibrium state characterized by the
temperatures $T_{\perp\alpha }$ and $T_{\Vert\alpha }$.

\section{Linear instability}

Before turning to the nonlinear regime, we briefly review the derivation of
the MI linear growth rate in the simplified framework provided by the drift
kinetic approximation which is only valid at scales large enough for FLR
effects to be subdominant.

Linearizing  Eq. (\ref{balance-nl}) about the background field $\mathbf{%
B_{0}}$ and equilibrium pressures $p_\perp^{(0)}$ and  $p_\|^{(0)}$, 
and considering  perturbations $\widetilde{\mathbf{B}}$ and 
$p_\perp^{(1)}\propto  e^{-i\omega t+i\mathbf{k\cdot r}}$, we get 
\begin{equation}
p_{\perp }^{(1)}+\frac{B_{0}\widetilde{B}_{z}}{4\pi }=-\frac{k_{z}^{2}}{%
k_{\perp }^{2}}\Big(1+\frac{\beta _{\perp }-\beta _{\Vert }}{2}\Big)\frac{%
B_{0}\widetilde{B}_{z}}{4\pi }.  \label{pressures-1}
\end{equation}%
Here, $p_{\perp }^{(1)}$ has to be calculated from the linearized drift kinetic
equation
\begin{equation}
\frac{\partial f_{\alpha }^{(1)}}{\partial t}+v_{\Vert }\frac{\partial
f_{\alpha }^{(1)}}{\partial z}+\Big [-\mu \frac{\partial \widetilde{B}_{z}}{%
\partial z}+\frac{e_{\alpha }}{m_{\alpha }}E_{\Vert }\Big]\frac{\partial
f_{\alpha }^{(0)}}{\partial v_{\Vert }}=0,  \label{kin-1}
\end{equation}%
where we assume each $f_{\alpha }^{(0)}$ to be a bi-Maxwellian distribution
function 
\begin{equation}
f_{\alpha }^{(0)}=A_{\alpha }\exp \Big [-\frac{v_{\parallel }^{2}}{%
v_{\parallel \alpha }^{2}}-\frac{\mu B_{0}m_{\alpha }}{T_{\perp \alpha }}%
\Big ],  \label{bi-maxwell}
\end{equation}%
with $A_{\alpha }~=~n_{0}m_{\alpha }/(2\pi \sqrt{\pi }v_{\parallel \alpha
}T_{\perp \alpha })$.

Equation (\ref{kin-1}) is solved in Fourier representation, as 
\begin{equation}
f_{\alpha }^{(1)}=-\frac{\mu \widetilde{B}_{z}+\frac{e_{\alpha }} {m_{\alpha
}}\phi }{\omega -k_{z}v_{\Vert }}k_{z}\frac{\partial f_{\alpha }^{(0)}}{
\partial v_{\Vert }}.  \label{kin-gen-1}
\end{equation}
The neutrality condition 
allows one to express the potential $\phi$ in terms of $\widetilde{B}
_{z}$. Indeed, assuming   $\displaystyle{%
\zeta ={\sqrt{\pi }\omega }/(|k_{z}|v_{\parallel i})\ll 1}$
(so that the contribution from the Landau pole is small),
\begin{equation}
\int f_{i}^{(1)}dv_{z}d\mu d\varphi =-\frac{n_{0}}{B_{0}T_{\parallel i}} %
\Big[ T_{\perp i}\frac{\widetilde{B}_{z}}{B_{0}}+e\phi \Big ] \Big [ 1+ 
i\zeta \Big ].  \nonumber
\end{equation}
Similarly, neglecting the electron
Landau resonance contribution because of the small mass ratio, 
\begin{equation}
\int f_{e}^{(1)}dv_{\|}d\mu d\varphi =-\frac{n_{0}}{B_{0}T_{\Vert e}}\Big[ %
T_{\perp e}\frac{\widetilde{B}_{z}}{B_{0}}-e\phi \Big].  \nonumber
\end{equation}
Consequently,
\begin{equation}
e\phi \approx \frac{T_{\perp i}}{1+\theta _{\parallel }}\Big[(\theta _{\perp
}-\theta _{\parallel })-\frac{\theta _{\parallel }(1+\theta _{\perp })}{
1+\theta _{\parallel }}i\zeta \Big]\frac{\widetilde{B}_{z}}{B_{0}}.
\label{phi-1}
\end{equation}%
We thus recover that for mirror modes, the parallel electric field
vanishes when the electrons are cold ($\theta_\perp = \theta_\| = 0$).
Interestingly, when $\theta _{\perp }=\theta _{\parallel }$, only the Landau
pole contributes to $\phi$.

It is now necessary to evaluate 
\begin{equation}
p_{\perp }^{(1)}=2\frac{\widetilde{B}_{z}}{B_{0}}p_{\perp
}^{(0)}+B_{0}^{2}\sum_{\alpha }m_{\alpha }\int \mu f_{\alpha }^{(1)}d\mu
dv_{\Vert }d\varphi .  \nonumber
\end{equation}%
Using 
\begin{eqnarray}
\int \frac{k_{z}v_{\Vert }}{\omega -k_{z}v_{\Vert }}f_{i}^{(0)}d\mu
dv_{\Vert }d\varphi  &=&-\frac{n_{0}}{B_{0}}(1+i\zeta )  \nonumber \\
\int \frac{k_{z}v_{\Vert }}{\omega -k_{z}v_{\Vert }}f_{e}^{(0)}d\mu
dv_{\Vert }d\varphi  &=&-\frac{n_{0}}{B_{0}},  \nonumber
\end{eqnarray}%
we get 
\begin{equation}
p_{\perp }^{(1)}=-\beta _{\perp }\frac{B_{0}^{2}}{4\pi }\Big[\frac{1}{\beta
_{\perp }}+\Gamma +\frac{T_{\perp i}}{T_{\Vert i}}\frac{i\zeta D}{2(1+\theta
_{\perp })}\Big]\frac{\widetilde{B}_{z}}{B_{0}}.  \nonumber
\end{equation}%
Substituting this expression into the linearized force balance equation
yields the linear instability growth rate given by Eq. (\ref{growth_rate}),
up to the FLR term which is not captured by the drift kinetic approximation.
Note that the growth rate given by Eq. (\ref{growth_rate}) is consistent
with the applicability condition $\gamma /|k_{z}|\ll v_{{\Vert }i}$ near
threshold ($\Gamma \ll 1$), as 
$k_{z}$ and   $(k_{z}/k_{\perp })^2$ scale like  $\Gamma $,  while 
$\gamma$ like  $\Gamma ^{2}$. 

\section{General pressure estimates}

As demonstrated in \cite{KPS07a,KPS07b}, the scalings resulting from the
linear theory near threshold imply an adiabaticity condition to leading order.
It is thus enough to consider the stationary kinetic equation 
\begin{equation}
v_{\Vert }\mathbf{b}\cdot \nabla f_{\alpha }-(\mathbf{b}\cdot \nabla )\left[
\mu B+\frac{e_{\alpha }}{m_{\alpha }}\phi \right] \frac{\partial f_{\alpha } 
}{\partial v_{\Vert }}=0.  \label{Vlasov_stat}
\end{equation}
It turns out that Eq. (\ref{Vlasov_stat}) is exactly solvable, the
general solution being an arbitrary function $f_{\alpha}=g_{\alpha }(\mu
,W_{\alpha })$ of the particle energy $\displaystyle{\ W_{\alpha }=
{v_{\Vert }^{2}}/{2}+\mu B+\frac{e_{\alpha }}{m_{\alpha }} \phi}$, and of $\mu$.
To find the function $g_{\alpha }(\mu ,W_{\alpha })$, we use the
adiabaticity argument which means that, to leading order, $g_{\alpha }$ as a
function of $\mu$ and $W_{\alpha }$ retains its form during
the evolution. Therefore, the function $g_{\alpha }(\mu ,W_{\alpha })$ is
found by matching with the initial distribution function $f_{\alpha }^{(0)}$
given by Eq. (\ref{bi-maxwell}) which corresponds to $\phi =0$ and $%
W_{\alpha }=\frac{v_{\Vert }^{2}}{2}+\mu B_{0}$. We get 
\begin{eqnarray}
&&g_{\alpha }(\mu ,W_{\alpha })=A_{\alpha }\exp \Big[ -\frac{v_{\parallel
}^{2}}{v_{\parallel \alpha }^{2}}-\frac{\mu B_{0}m_{\alpha }}{T_{\perp
\alpha }}\Big]  \nonumber \\
&&\quad=A_{\alpha }\exp \Big [ -\frac{2W_{\alpha }} {v_{\parallel \alpha}^{2}}+
\mu B_{0}m_{\alpha }\Big( \frac{1}{T_{\parallel \alpha }}-\frac{1}{
T_{\perp \alpha }}\Big) \Big] .  \label{g-alpha}
\end{eqnarray}
Thus, $g_{\alpha }(\mu ,W_{\alpha })$ is a Boltzmann distribution function
with respect to $W_{\alpha }$ but, at fixed $W_{\alpha }$, it displays an
exponential growth relatively to $\mu $ if $T_{\perp \alpha }>$ $%
T_{\parallel \alpha }$. This effect can however be compensated by the
dependence of $W_{\alpha }$ in $\mu$. This means that only a fraction of the
phase space $(\mu ,W_{\alpha })$ is accessible, a property possibly related
with the existence of trapped and untrapped particles.

Note that expanding Eq. (\ref{g-alpha}) relatively to $\widetilde{B}_{z}/B_0$
and $e\phi ^{(1)}/T_{\perp i}$ reproduces the first order contribution to the
distribution function given by Eq. (\ref{kin-gen-1}) with $\omega=0$, and also
the second order correction found in \cite{KPS07a,KPS07b} 
in the case of cold electrons. It should be
emphasized that Eq. (\ref{g-alpha}) only assumes adiabaticity and remains
valid for finite perturbations.

The function $g_{\alpha }$ can also be rewritten in terms of $v_{\Vert }$, $%
v_{\perp }$ and $\phi $ as 
\begin{eqnarray}
&&g_{\alpha }=A_{\alpha }\exp \Big[-\frac{m_{\alpha }v_{\Vert }^{2}}{
2T_{\parallel \alpha }}-\frac{e_{\alpha }\phi }{T_{\parallel \alpha }}\Big] %
\times  \nonumber \\
&&\qquad \exp \left\{ -\frac{m_{\alpha }v_{\perp }^{2}}{2T_{\perp \alpha }} %
\Big (\frac{T_{\perp \alpha }}{T_{\parallel \alpha }}-\frac{B_{0}}{B}\Big [ 
\frac{T_{\perp \alpha }}{T_{\parallel \alpha }}-1\Big ]\Big )\right\} , 
\nonumber
\end{eqnarray}%
which can be viewed as the bi-Maxwellian distribution function with the
renormalized transverse temperature 
\begin{equation}
T_{\perp \alpha }^{(eff)}=T_{\perp \alpha }\left[ \frac{T_{\perp \alpha }}{
T_{\parallel \alpha }}-\frac{B_{0}}{B}\Big(\frac{T_{\perp \alpha }}{
T_{\parallel \alpha }}-1\Big)\right] ^{-1}.  \nonumber
\end{equation}%
Note the Boltzmann factor $\exp {-[e_{\alpha }\phi /T_{\parallel \alpha }]}$
in the expression of $g_{\alpha }$. For cold electrons, the ion distribution
function was obtained in \cite{Const02} by assuming that it 
remains bi-Maxwellian, and owing to the invariance of the kinetic energy and
of the magnetic moment. This estimate, obtained by neglecting both time
dependency (and consequently  Landau resonance) and finite Larmor radius
corrections, reproduces the closure condition given in \cite{PRS06}.

After rewriting Eq. (\ref{g-alpha}) in the form 
\begin{equation}
g_{\alpha }=A_{\alpha }\exp \Big[-\frac{e_{\alpha }\phi }{T_{\parallel
\alpha }}-\frac{v_{\Vert }^{2}}{v_{\parallel \alpha }^{2}}-\frac{\mu
B_{0}m_{\alpha }}{T_{\perp \alpha }}\Big(1+\frac{T_{\perp \alpha }}{
T_{\parallel \alpha }}\frac{B-B_{0}}{B_{0}}\Big)\Big],  \nonumber
\end{equation}
the quasi-neutrality condition gives 
\begin{eqnarray}
&&\left( 1+\frac{T_{\perp i}}{T_{\parallel i}}\frac{B-B_{0}}{B_{0}}\right)
^{-1}\exp \left( -\frac{e\phi }{T_{\parallel i}}\right) =  \nonumber \\
&&\left( 1+\frac{T_{\perp e}}{T_{\parallel e}}\frac{B-B_{0}}{B_{0}}\right)
^{-1}\exp \left( \frac{e\phi }{T_{\parallel e}}\right)  \nonumber
\end{eqnarray}
or 
\begin{eqnarray}
&&e\phi =(T_{\parallel i}^{-1}+T_{\parallel e}^{-1})^{-1}\times  \nonumber \\
&&\log \left[ \left( 1+\frac{T_{\perp e}}{T_{\parallel e}}\frac{B-B_{0}}{
B_{0}}\right) \left( 1+\frac{T_{\perp i}}{T_{\parallel i}}\frac{B-B_{0}}{
B_{0}}\right) ^{-1}\right] .  \label{potential}
\end{eqnarray}%
Interestingly, the electron density (and thus also that of the ions)
\begin{equation}
n_{e}=n_{0}\frac{B}{B_{0}}\left( 1+\frac{T_{\perp e}}{T_{\parallel e}}\frac{
B-B_{0}}{B_{0}}\right) ^{-1}\exp \left[ \frac{e\phi }{T_{\parallel e}}\right]
\nonumber
\end{equation}%
has the usual Boltzmann factor $\exp \left[ e\phi /T_{\parallel e}\right] $
and also an algebraic prefactor depending on the magnetic field $B$. In the
case of isotropic electron temperature ($T_{\perp e}=T_{\parallel e}\equiv
T_{e}$), the electron density has the usual Boltzmann form $n_{e}=n_{0}\exp %
\left[ e\phi /T_{e}\right] $.

Equation (\ref{potential}) shows that the potential vanishes in two
cases: for cold electrons and also when electron and ion temperature anisotropies 
$a_e$ and $a_i$ (with $a_\alpha = T_{\perp \alpha} /T_{\|\alpha}$) 
are equal, a case considered in the linear theory of the
mirror instability \cite{Stix62,hasegawa,Hall79}.

In order  to evaluate explicitly the perpendicular pressure for each species 
\begin{eqnarray}
&&p_{\perp \alpha }=m_{\alpha }B^{2}\int \mu g_{\alpha }d\mu dv_{\Vert
}d\varphi   \nonumber \\
&&\ \ =n_{0}T_{\perp \alpha }\frac{B^{2}}{B_{0}^{2}}\Big(1+\frac{T_{\perp
\alpha }}{T_{\parallel \alpha }}\frac{B-B_{0}}{B_{0}}\Big)^{-2}\exp \Big(-%
\frac{e_{\alpha }\phi }{T_{\parallel \alpha }}\Big),  \nonumber
\end{eqnarray}%
where $e\phi $ is given by Eq. (\ref{potential}), it is convenient to
introduce the functions 
\begin{eqnarray}
S_{\perp i}(u) &=&\left( \frac{1+u}{1+a_{i}u}\right) ^{2}\left( \frac{%
1+a_{i}u}{1+a_{e}u}\right) ^{c_{i}}  \label{Sperpi} \\
S_{\perp e}(u) &=&\left( \frac{1+u}{1+a_{e}u}\right) ^{2}\left( \frac{%
1+a_{e}u}{1+a_{i}u}\right) ^{c_{e}},  \label{Sperpe}
\end{eqnarray}%
with the notations $u=(B-B_{0})/B_{0}$ and  $c_{\alpha }=T_{\parallel \alpha
}^{-1}/(T_{\parallel i}^{-1}+T_{\parallel e}^{-1})$. The two latter
functions transform one into the other by exchanging the subscripts $i$ and $%
e$. The ion and electron perpendicular pressures are then written as $%
p_{\perp \alpha }=n_{0}T_{\perp \alpha }S_{\perp \alpha }(u)$. In the
special case of cold electrons, 
\begin{equation}
p_{\perp }=n_{0}T_{\perp i}\frac{B^{2}}{B_{0}^{2}}\Big(1+\frac{T_{\perp i}}{%
T_{\parallel i}}\frac{B-B_{0}}{B_{0}}\Big)^{-2},  \label{p_perb-B}
\end{equation}%
which is algebraic relatively to $B$. From  this expression 
as well as from the general formula for  $p_{\perp }=p_{\perp i}+p_{\perp e}$
given by Eqs. (\ref{Sperpi}) and (\ref{Sperpe}) it follows that the perpendicular
and  magnetic pressures are anticorrelated. When $B$  increases (decreases), 
the ratio of the perpendicular to the magnetic 
pressure, i.e. the local $\beta _{\perp }$,
decreases (increases), which  corresponds to a reduction (an increase)
of the distance to threshold. This implies that the instability 
cannot  saturate at  small amplitudes. 

Similarly, for the parallel pressure, we have 
\begin{equation}
p_{_{\parallel \alpha }}=n_{0}T_{_{\parallel \alpha }}\frac{B}{B_{0}}\Big(1+%
\frac{T_{\perp \alpha }}{T_{\parallel \alpha }}\frac{B-B_{0}}{B_{0}}\Big)%
^{-1}\exp \Big(-\frac{e_{\alpha }\phi }{T_{\parallel \alpha }}\Big). 
\nonumber
\end{equation}%
that rewrites $p_{\Vert \alpha} =n_{0}T_{\Vert \alpha}S_{\Vert \alpha}(u)$ with 
\begin{eqnarray}
S_{\| i}(u) &=&\left (\frac{1+u}{1+a _{i}u}\right ) \left (\frac{1+a _{i}u}{%
1+a _{e}u}\right)^{c_i}  \label{Sparali} \\
S_{\| e}(u) &=&\left(\frac{1+u}{1+a _{e}u}\right) \left (\frac{1+a _{e}u}{
1+a _{i}u}\right)^{c_e}.  \label{Sparale}
\end{eqnarray}%

\section{ The weakly nonlinear regime}

As it follows from Eq. (\ref{pressures-1}), in the linear regime near 
threshold, the fluctuations of perpendicular and magnetic
pressures  almost compensate each other. 
In the weakly nonlinear regime, the second order
correction to the total (perpendicular plus magnetic) pressure is thus relevant
and leads to a local shift of $\Gamma$. To find this correction, we consider  the
expansions of  the perpendicular pressures of the ions and electrons in the 
$u$ variable. Because of the symmetry between the functions $S_{\perp
i}(u)$ and $S_{\perp e}(u)$, it is enough to consider the expansion 
\begin{eqnarray}
S_{\perp i}(u) &=&1+u\Big (2-2a _{i}-c_{i}(a _{e}-a _{i})\Big ) \nonumber \\
&&+u^{2}\Big [c_{i}\Big (a _{e}a _{i}-a _{i}^{2}+\frac{1}{2}(a_{e}-a _{i})^{2}\Big)  
\nonumber \\
&&-4a _{i}+3a _{i}^{2}+\frac{1}{2}c_{i}^{2}(a _{e}-a
_{i})^{2}-2c_{i}(a _{e}-a _{i})  \nonumber \\
&&+2\alpha _{i}c_{i}(a _{e}-a _{i})+1\Big ]+O\left( u^{3}\right)  \nonumber
\end{eqnarray}%
As a result, the second order contributions to the  perpendicular ion  pressure
is  given by 
\begin{eqnarray}
&&p_{i\perp }^{(2)}=n_{0}T_{\perp i}\Big [%
3a _{i}^{2}-4a _{i}+1+c_{i}(a _{e}-a _{i})  \nonumber \\
&&\qquad \times \Big(\frac{1}{2}(c_{i}+1)(a _{e}-a _{i})-2+3a _{i}%
\Big)\Big ]u^2,  \nonumber 
\end{eqnarray}%
with an analogous formula for the perpendicular electron pressure, 
obtained by exchanging the $i$ and $e$ indices.
   Furthermore,  the threshold condition rewrites
\begin{eqnarray}
&&\frac{B_{0}^{2}}{4\pi }+n_{0}\Big \{T_{\perp i}\left[ 2-2a
_{i}-c_{i}\left( a _{e}-a _{i}\right) \right]   \nonumber \\
&&\qquad +T_{\perp e}\left[ 2-2a _{e}+c_{e}\left( a _{e}-a
_{i}\right) \right] \Big \}=0.  \label{new-threshold}
\end{eqnarray}%
The quadratic contributions to the pressure balance (\ref{balance-nl}), 
originating from $p_{i\perp }^{(2)}+p_{e\perp
}^{(2)}+\left( B-B_{0}\right) ^{2}/(8\pi )$, are collected in a  term 
 $\Lambda \left( \frac{B-B_{0}}{B_{0}}\right) ^{2}$ with 
\begin{eqnarray}
\Lambda  &=&n_{0}\Big \{T_{\perp i}\Big (3a _{i}^{2}-4a _{i}+1 
\nonumber \\
&&+c_{i}(a _{e}-a _{i})\Big [\frac{1}{2}(1+c_{i})(a _{e}-a
_{i})-2+3a _{i}\Big ]\Big )  \nonumber \\
&&+T_{\perp e}\Big (3a _{e}^{2}-4a _{e}+1+c_{e}(a _{e}-a _{i}) 
\nonumber \\
&&\times \Big [\frac{1}{2}(1+c_{e})(a _{e}-a _{i})+2-3a _{e}\Big ]%
\Big )\Big \}+\frac{B_{0}^{2}}{8\pi }.  \label{eqlambda}
\end{eqnarray}%
The value $\Lambda _{c}$ of  $\Lambda $ at
threshold is obtained  by expressing ${B_{0}^{2}}/{8\pi }$ by
means of Eq.  (\ref{new-threshold}), which gives 
\begin{eqnarray}
\Lambda _{c} &=&n_{0}\Big \{T_{\perp i}\Big [3a _{i}^{2}-4a _{i}+1 
\nonumber \\
&&+c_{i}(a _{e}-a _{i})\Big (\frac{1}{2}(1+c_{i})(a _{e}-a
_{i})-2+3a _{i}\Big )  \nonumber \\
&&-\frac{1}{2}\Big (2-2a _{i}-c_{i}(a _{e}-a _{i})\Big )\Big ] 
\nonumber \\
&&+T_{\perp e}\Big [3a _{e}^{2}-4a _{e}+1+c_{e}(a _{e}-a _{i}) 
\nonumber \\
&&\times \Big (\frac{1}{2}(1+c_{e})(a _{e}-a _{i})+2-3a _{e}] 
\nonumber \\
&&-\frac{1}{2}\Big (2-2a _{e}+c_{e}(a _{e}-a _{i})\Big )\Big ]\Big \}. \nonumber
\end{eqnarray}%
After some algebra, defining $\lambda _{c}=\Lambda _{c}/(n_{0}T_{\perp i})$, one gets 
\begin{eqnarray}
&&\frac{\lambda _{c}}{\alpha _{i}}=\frac{T_{\perp i}}{T_{\parallel i}}\Big [%
3+3\frac{\theta _{\perp }^{3}}{\theta _{\parallel }^{2}}-\frac{1}{2}\frac{%
\left( \theta _{\perp }-\theta _{\parallel }\right) ^{2}}{\theta _{\parallel
}^{2}\left( 1+\theta _{\parallel }\right) ^{2}}  \nonumber \\
&&\qquad \times \left( 4\theta _{\perp }+4\theta _{\parallel
}^{2}+\allowbreak 5\left( \theta _{\perp }+1\right) \theta _{\parallel
}\right) \Big]  \nonumber \\
&&\qquad -\frac{3}{2\theta _{\parallel }\left( 1+\theta _{\parallel }\right) 
}\left[ \left( \theta _{\perp }+\theta _{\parallel }\right) ^{2}+2\theta
_{\parallel }(1+\theta _{\perp }^{2})\right]. \label{eqlambda_c}
\end{eqnarray} %

Proceeding as in \cite{KPS07a}, retaining the contribution of the 
above  quadratic terms to the pressure balance, leads one to supplement 
a nonlinear contribution to  Eq. (\ref{growth_rate}) that becomes 
\begin{eqnarray}
&&\frac{\partial u}{\partial t}=\frac{2}{\sqrt{\pi }}\frac{T_{\Vert i}}{%
T_{\perp i}}\frac{v_{\Vert i}}{D}{\widehat{{\cal K}_z }}
\Big \{ \Gamma u -\frac{\chi}{\beta _{\perp }}(\Delta_\perp )^{-1}\partial _{zz}u \nonumber \\
&& +\frac{3}{4}\Big (\frac{T_{\perp i}}{T_{\Vert i}}-1\Big )%
\frac{1+F}{1+\theta _{\perp }}r_{L}^{2}\Delta _{\perp }u
-\frac{\lambda _{c}}{2(1+\theta _{\perp })}u^2 \Big \}\label{NL}
\end{eqnarray}%
Here the integral operator ${\widehat{{\cal K}}_z}$ reduces in Fourier representation
to $|k_z|$ and  $\chi = 1+\frac{\beta_{\perp }-\beta _{\Vert }}{2}$. Furthermore, within the present
approximation, $u$ coincides with ${\widetilde B}_z/B_0$.

Equation (\ref{NL}) extends the result of \cite{KPS07a, Calif08} valid for 
cold electrons. As in the latter case,  this equation is a gradient type equation,
\[
\frac{\partial u}{\partial t}=-{\widehat{{\cal K}_z }}\frac{\delta F}{\delta u},
\]
for which the free
energy (written in dimensionless variables)
\[
F=\int \left \{  \frac 12 \left [ -\Gamma u^2 +(\partial_z u)^2 +u\Delta_\perp ^{-1}\partial _{zz}u\right ] +\frac 13\lambda _{c}u^3\right \} d{\bf r}
\] 
is  unbounded  from below due to the integral $\int \lambda _{c}u^3 d{\bf r}$.
This leads to a blow-up behavior, associated with a subcritical bifurcation \cite%
{KPS07a}, \cite{KPS07b}. Saturation at large values of the amplitude 
and formation of stationary structures requires additional nonlinear
effects such as the influence of resonant particles on the nonlinear coupling \cite{PSKH09}.
Equation (\ref{NL}) that does not include saturation processes is not suitable to address the 
question of the reduction of the temperature anisotropy by the development of the mirror
instability mentioned in \cite{VedenovSagdeev}. This effect is reproduced by the 
quasi-linear theory \cite{SS64}, and was also studied in the context of the 
so-called FLR-Landau fluid model
that, like the present asymptotics, retains a linear description of the Landau resonance and 
of FLR effects, but includes all the hydrodynamic nonlinearities and does not a priori
prescribe a pressure balance condition. It was  observed in this case 
that during the saturation phase, the mean temperatures rapidly evolve in a way as to   
reduce the distance to threshold \cite{Borgogno07}.

As demonstrated in \cite{KPS07a,KPS07b}, 
the sign of the nonlinear coupling $\lambda _{c}$ 
defines the type of the mirror structures, namely holes ($\lambda _{c}>0$) or 
humps ($\lambda _{c}<0$), near threshold. This sign is strongly dependent on the
equilibium distribution function \cite{HKP09} . It is nevertheless of interest to consider the case
where both  ions and electrons have a bi-Maxwellian distribution function.
It turns out that the sign of $\lambda _{c}$
can then be determined analytically in a few special cases.

\noindent
{ \it (i) Limit $\protect\theta_\| \ll \protect\theta_\perp$:}
\begin{equation}
\frac{\Lambda _{c}}{n_{0}T_{\perp i}a _{i}}=\frac{\theta _{\perp }^{2}}{%
\theta _{\parallel }}\left( \frac{T_{\perp e}}{T_{\parallel e}}-\frac{3}{2}%
\right) >0.\nonumber
\end{equation}

\noindent
{\it (ii) Equal anisotropies ($\protect\theta _{\perp}=\protect\theta_{\parallel }$)}
\begin{eqnarray}
\Lambda _{c} &=&n_{0}(T_{\perp i}+T_{\perp e})\left( 3a ^{2}-4a
+1\right)   \nonumber \\
&&-n_{0}(T_{\perp i}+T_{\perp e})\left( 1-a \right) =3a \frac{B_{0}^{2}%
}{8\pi }>0.\nonumber
\end{eqnarray}

\noindent
{\it (iii) Isotropic electron temperature:}
The coefficient
$\Lambda _{c}$ can be rewritten in the form 
\begin{eqnarray}
\Lambda _{c} &=&n_{0}(a _{i}-1)\{T_{\perp i}\Big((3a _{i}-1)  \nonumber
\\
&&+c_{i}\Big[\frac{1}{2}(1+c_{i})\left( \alpha _{i}-1\right) +2-3a _{i}%
\Big]\Big )  \nonumber \\
&&+T_{e}c_{e}\Big[\frac{1}{2}\left( 1+c_{e}\right) \left( a _{i}-1\right)
+1\Big]\}+\frac{B_{0}^{2}}{8\pi }.  \nonumber
\end{eqnarray}%
   Furthermore, at  threshold, 
\begin{equation}
\frac{1}{2}n_{0}(a _{i}-1)\left[ T_{\perp i}\left( 2-c_{i}\right)
+T_{\perp e}c_{e}\right] =\frac{B_{0}^{2}}{8\pi }>0.\nonumber
\end{equation}%
Hence, we simultaneously have two inequalities $a _{i}>1$ and $T_{\perp
e}c_{e}>T_{\perp i}(c_{i}-2)$. Therefore, 
\begin{eqnarray}
\Lambda _{c} &=&n_{0}(a _{i}-1)\Big \{T_{\perp i}\Big((3a _{i}-1) 
\nonumber \\
&&+c_{i}\Big [\frac{1}{2}(1+c_{i})\Big(a _{i}-1\Big)+2-3a _{i}\Big]%
\Big)  \nonumber \\
&&+T_{e}c_{e}\Big[\frac{1}{2}(1+c_{e})(a _{i}-1)+1\Big ]\}  \nonumber \\
&&+\frac{1}{2}n_{0}(a _{i}-1)\left[ T_{\perp i}\left( 2-c_{i}\right)
+T_{\perp e}c_{e}\right]   \nonumber \\
&=&n_{0}(a _{i}-1)\Big \{T_{\perp i}\Big(3a _{i}(1-c_{i})  \nonumber \\
&&+c_{i}\Big[\frac{1}{2}(1+c_{i})\left( a _{i}-1\right) +\frac{3}{2}\Big]%
\Big)  \nonumber \\
&&+T_{e}c_{e}\Big[2+\frac{1}{2}\left( 1+c_{e}\right) \left( a
_{i}-1\right) \Big]\Big \},\nonumber
\end{eqnarray}%
which is positive, because $1-c_{i}= c_{e}=
(1+\theta _{\parallel })^{-1}>0$ and $a _{i}>1$.

\begin{figure}[t]
\centerline{
\includegraphics[width=0.46\textwidth]{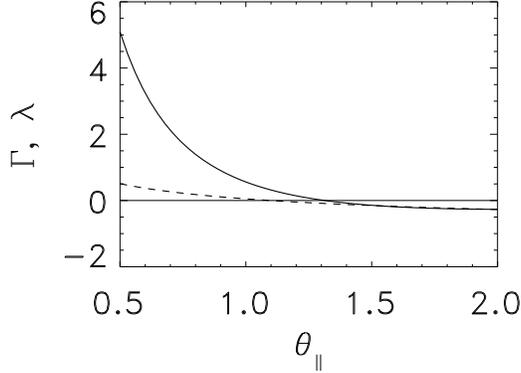}
}
\caption{Fig. 1. Variation with $\protect\theta_\|$ of the distance to
threshold $\Gamma$ given by Eq. (\protect\ref{newthreshold}) (dashed line)
and of the normalized nonlinear coupling coefficient $\protect\lambda$
(solid line) evaluated from Eq. (\protect\ref{eqlambda}) for $\protect\theta_{\perp}=1$
, $\protect a_i=1.1$ and $\protect\beta_{\perp i} = 10$.}
\label{fig1}
\end{figure}

\begin{figure}[t]
\centerline{
\includegraphics[width=0.46\textwidth]{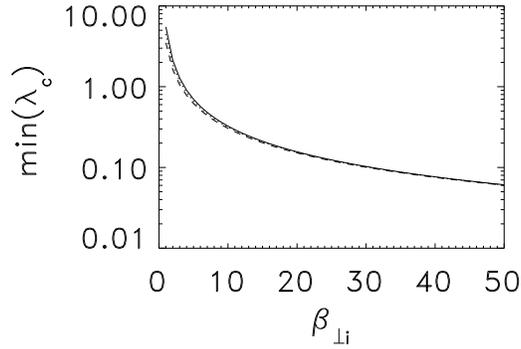}
}
\caption{Fig. 2. Variation with $\protect\beta_{\perp i}$ of the minimum $%
\mathrm{min}\,( \protect\lambda_c)$ of the normalized nonlinear coupling
coefficient taken in an interval of values of $a_p$ between $0$ and $a_{p1}(%
\protect\beta_{\perp i})$, defined such that the threshold is obtained for a
value of $\protect\theta_\|$ equal to $100$, for $\protect\theta_\perp= 0.2$
(solid line), $\protect\theta_\perp=1$ (dotted line) and $\protect\theta%
_\perp=5$ (dashed line).}
\label{fig2}
\end{figure}

\noindent
{\it (iv)  More general conditions:}
A numerical approach was used in this case. Figure  1 displays, for typical values 
of the parameters (taken
here as $\theta _{\perp }=1$, $a _{i}=1.1$ and $\beta _{\perp i}=10$), the
distance to threshold $\Gamma $ (dashed line) given by Eq. (\ref
{newthreshold}) and the non-dimensional nonlinear coupling coefficient $%
\lambda =\Lambda /(n_{0}T_{\perp i})$ (solid line), where $\Lambda $ is given by
Eq. (\ref{eqlambda}), as a function of $\theta _{\Vert }$. This graph is
typical of the general behavior of these functions and shows that they are
both decreasing as $\theta _{\Vert }$ increases, with $\lambda $ possibly
reaching negative values, but only below threshold. In order to show that
the value $\lambda _{c}$, given by Eq. (\ref{eqlambda_c}), of $\lambda $ 
at threshold is positive in a wider range of parameters, we display in Fig.
2, as a function of $\beta _{\perp i}$ for $\theta _{\perp }=0.2$ (solid
line), $\theta _{\perp }=1$ (dotted line) and $\theta _{\perp }=5$ (dashed
line), the quantity $\mathrm{{min}\,(\lambda _{c})}$ obtained after
minimizing $\lambda _{c}$ in an interval of values of $a_{p}$ between $0$
and $a_{p1}(\beta _{\perp i})$. The latter quantity is arbitrarily defined
such that the threshold is obtained for a value of $\theta _{\Vert }$ equal
to $100$. This graph shows that $\mathrm{min}(\lambda _{c})$ varies little
with $\theta _{\perp }$ but is very sensitive to $\beta _{\perp i}$. As the
latter parameter is increased, $\mathrm{{min}\,(\lambda _{c})}$ decreases
but remains always positive. Although 
this numerical observation is  
not a rigorous proof, it convincingly shows that $\Lambda>0 $
in the parameter range  of physical interest.

\section{Stationary nonlinear structures}

Substituting the explicit expressions of the gyrotropic pressures
in terms of the magnetic field amplitude
given in the Section IV, within  the equation
for the balance of forces 
\begin{eqnarray}
&&-\nabla \Big( p_{\perp }+\frac{B^{2}}{8\pi }\Big) +\Big[ 1+\frac{4\pi } {%
B^{2}}( p_{\perp }-p_{\Vert }) \Big] \frac{(\mathbf{B}\cdot \nabla )\mathbf{B}%
}{4\pi }  \nonumber \\
&&\qquad + \mathbf{B}(\mathbf{B}\cdot\nabla) \Big (\frac{p_\perp -p_\|}{B^2} \Big ) \, 
 =0, \label{equil_noFLR}
\end{eqnarray}
leads to a closed system that seems overdetermined due to the divergenceless condition 
$\nabla \cdot {\bf B}=0$. In fact, it can be checked, after some algebra
using the explicit expressions  (\ref{Sperpi},\ref{Sperpe}) and (\ref{Sparali},\ref{Sparale}),   
that the projection of Eq. 
(\ref{equil_noFLR}) on the magnetic field vanishes identically, thus reducing the 
system to three equations for three unknowns.
These equations  can be useful  for finding, possibly numerically,  
stationary profiles of three-dimensional finite-amplitude 
stationary mirror structures. Note that Eq. (\ref{equil_noFLR}) differs from the 
Grad-Shafranov equation  \cite{grad,shafra} in  that the parallel and perpendicular
pressures are here prescribed functions of the magnetic field amplitude.
A main issue concerns the existence of 
stable subcritical solutions, a question that is beyond the scope of this 
letter and will be addressed in forthcoming works. 
Such structures are reported by satellite observations \cite{SLD07,Genot09} 
and are also expected from the subcritical character of the
mirror instability \cite{KPS07b}. Equilibrium solutions
were computed  in one-space dimension in 
\cite{PRS06}, where they lead to discontinuous profiles. Their regularization would 
require that FLR corrections be retained. These  additional contributions are known 
from the linear kinetic theory but their extension to the finite-amplitude
case remains a challenging problem.

This work was supported by the CNRS PICS programme 6073 and RFBR grant
12-02-91062-CNRS\_a. T.P. and P.L.S. benefited from support from INSU-CNRS
 Programme National PNST. The work of
E.K. was also supported by the RAS Presidium Program "Fundamental problems
of nonlinear dynamics in mathematical and physical sciences",  Grant NSh
7550.2006.2 and by the French Minist\`ere de l'Enseignement Sup\'erieur 
et de la Recherche.



\end{document}